# Laminar flow characterization using low-field magnetic resonance techniques

Jiangfeng Guo (郭江峰), Michael M.B. Ross, Benedict Newling, Maggie Lawrence, Bruce J. Balcom[*]

UNB MRI Centre, Department of Physics, University of New Brunswick, Fredericton, New Brunswick, E3B 5A3, Canada

[*] Corresponding author, Email address: bjb@unb.ca

**ABSTRACT:** Laminar flow velocity profiles depend heavily on fluid rheology. Developing methods of laminar flow characterization, based on low-field magnetic resonance (MR), contribute to the widespread industrial application of the MR technique in rheology. In this paper, we outline the design of a low-cost, palm-sized permanent magnet with a $^1$H resonance frequency of 20.48 MHz to measure laminar flow. The magnet consists of two disk magnets, which were each tilted at an angle of 1° from an edge separation of 1.4 cm to generate a constant gradient, 65 G/cm, in the direction of flow. Subsequently, a series of process methods, for MR measurements, were proposed to characterize Newtonian and non-Newtonian fluid flows in a pipe, including phase-based method, magnitude-based method, and a velocity spectrum method. The accuracy of the proposed methods was validated by simulations, and experiments in Poiseuille flow and shear-thinning flow with the designed magnet. The new velocity profile methods proposed are advantageous because the MR hardware and measurement methods are simple and will result in a portable instrument. Although the governing equations are complicated, the data analysis is straightforward.

# 1. INTRODUCTION

Laminar flow, in fluid dynamics, is characterized by fluid micro elements flowing in parallel layers [1]. The characterization of laminar flow, including measurements of average velocity and velocity profile, is of considerable value in chemical and allied processing industries [2–6]. Various fluids exhibit different flow behaviours under laminar conditions, dependent on fluid rheological properties [7–13]. Laminar flow characterization is therefore helpful to characterize rheological properties.

Magnetic resonance/magnetic resonance imaging (MR/MRI) is attractive for flow measurements because of its non-invasive capabilities for measuring optically opaque objects [13–15]. Multiple MR- and MRI-based methods have been reported to characterize fluid flow. MRI-based methods for measuring flows are based on the application of magnetic field gradients, including frequency-, phase-, and motion-encoding gradients, to yield quantitative information about velocity distributions of the flowing fluid [14–20]. There are also some modified MRI-based methods that only use one type of gradient [21–24]. MRI-based methods, resolving flow velocity profiles, have been used to measure various types of flows, for example laminar flow [19, 20, 25], turbulence [26–29], and flow in porous media [30–32]. Unfortunately, these measurements are predominantly performed on laboratory research instruments. The chief challenges to MRI-based methods, for widespread industrial application, are the expense of the superconducting equipment and the demand of the high-performance gradient systems. MR-based methods for flow measurements do not need complicated equipment, a permanent magnet with a static magnetic field gradient is sufficient. MR methods therefore have the prospect of industrial application and great potential in characterizing fluid flow.

MR-based methods for characterizing flow are based on the effect of flow on the MR signal. Suryan (1951) measured MR signals of flowing water in a U-tube between the pole pieces of a magnet at 20 MHz, and reported the continuous wave MR signal increased as the partially saturated spins were replaced by unsaturated flowing spins [33]. Singer (1959) exploited this principle to demonstrate *in vivo* flow measurements [34]. Hirschel and Libello (1962) showed the steady state MR signal is a function of fluid velocity in the presence of flow [35]. Arnold and Burkhart (1965) employed a spin echo to study the influence of flow on MR signal under laminar flow conditions [36]. Stejkal (1965), Grover and Singer (1971), and Hayward et al. (1972) extended this work using a pulsed field gradient technique [37–39].

Since the effect of flow on the MR signal was first studied, multiple MR-based methods for characterizing flows have been reported. These methods can be classified into two main categories: (1) net phase accumulation-based techniques [38–40] and (2) magnitude-based time-of-flight techniques [41–44]. Net phase accumulation-based techniques rely on the application of a constant or pulsed magnetic field gradient in the direction of flow. The phase shift of the signal detected is proportional to the average velocity component in the direction of the gradient. For example, Song et al. (2005) employed the Multiple Modulation Multiple Echoes (MMME) technique to measure fluid flow with a static magnetic field gradient [40]. A series of coherence pathways were generated by the MMME technique, and each of them exhibits a phase shift dependent on average velocity. Magnitude-based time-of-flight techniques are based on the variation of signal magnitude proportional to the quantity of excited spins in the detector, related to flow velocity. These techniques do not require the use of any magnetic field gradients. It is therefore very popular to employ this technique in low-cost low-field MR spectrometers. Beyond the two basic kinds of MR-based methods, O'Neill et al. invented an Earth's field magnetic resonance flow meter to measure the velocity probability distribution and $T_1$-velocity correlation probability distribution of multiphase flow [45–47].

From the descriptions of existing low-cost MR-based methods, it can be found these methods focus on the average velocity of fluid flow, which is insufficient to support the study of fluid rheology. The flow behaviour index is an important parameter in fluid rheology, which has a direct impact on the flow velocity profile under laminar conditions. Determination of the velocity profile is, therefore, helpful to study fluid rheology. In this paper, to make possible widespread industrial application of the MR technique in fluid rheology, we designed a low-cost, low-field, palm-sized permanent magnet with a flow-directed constant magnetic field gradient. Furthermore, we proposed corresponding MR-based methods to characterize laminar flow in a pipe, including average velocity and velocity profile, based on first odd echo Carr-Purcell-Meiboom-Gill (CPMG) MR measurements. The proposed methods were verified by simulations and flow experiments on the designed magnet.

## 2. METHODOLOGY

2.1. Equipment used

2.1.1 Sensor design and hardware

A magnet constructed with a separation between two disk magnets, as a function of distance along the symmetry axis, will intuitively lead to a magnetic field gradient directed along the symmetry axis. For the purposes of flow sensitization, this idea works remarkably well. Garwin and Reich [48] made a conceptually similar field modification with an aluminum plate added to an electromagnet for the purposes of diffusion sensitization in very early published MR measurement.

The optimal separation and tilt angle for a desired constant gradient of 60 G/cm between the two N52 NdFeB K&J Magnetics (Pipersville, PA) disk magnets of 5.1 cm diameter and 1.3 cm thickness was determined via CST Studio Suite (Providence, RI) simulation. In the simulation, each disk magnet was tilted at an angle of 1° from an edge separation of 1.4 cm between the magnets. 60 G/cm with this geometry was judged to be near ideal for the flow measurements envisaged. Fig. 1(a) depicts two disk magnets, each rotated by 1° about the *y* axis.

A 6 × 6 × 4 cm casing fabricated from Garolite G-10 (McMaster Carr, Elmhurst, IL) was machined to house the magnets. The casing was divided into two separate pieces, where each piece had a slot into which a magnet could be placed. Each slot was machined to permit the 1˚ tilt relative to the symmetry axis. The casing included a 1.2 cm diameter cylindrical hole through the shell, along the direction of the imposed gradient, to permit the placement of glass tubing to support the flow. A 4-turn solenoidal RF coil with 1.0 cm inner diameter was formed around a glass pipe, and was capacitively matched to 50 Ω. The RF coil, fabricated from 0.8 mm diameter copper wire, was centered in the Proteus (PROTon Embedded sUbmersible Sensor) magnet [49]. The interior and exterior of the Proteus magnet was wrapped with 0.2 mm copper tape to limit external RF interference and suppress acoustic ringing. Fig. 1(b) shows a photo of the Proteus magnet.

Magnetic field plots of the sensitive spot in the Proteus magnet were acquired with a LakeShore 460 3-Channel Gaussmeter (Westerville, OH) connected to a BiSlide Positioning System and VXM Stepping Motor Controller (Velmex Inc., Bloomfield, NY). Magnetic field data was read and processed through a custom MATLAB script (Mathworks, Natick, MA).

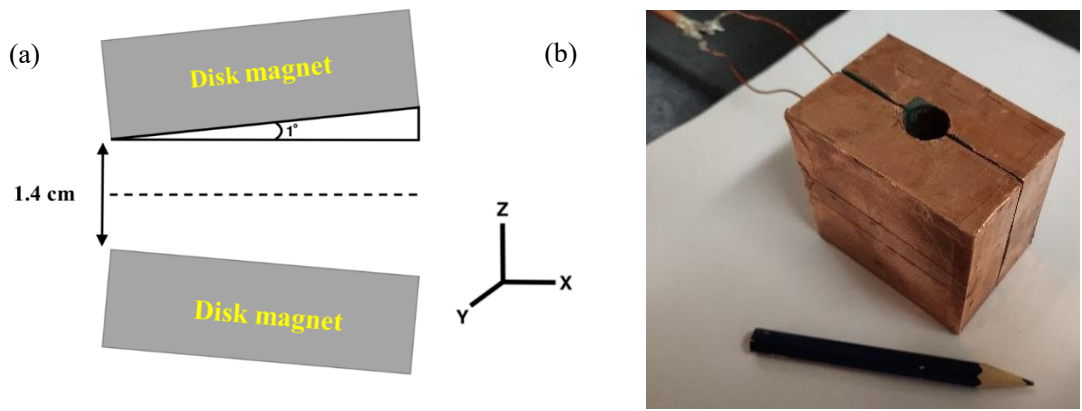

Fig. 1. (a) Diagram of the tilted disk magnets and (b) photo of the Proteus magnet. The two disk magnets are 5.1 cm in diameter and 1.3 cm in thickness. They were separated and tilted by 1° to generate a constant magnetic field gradient directed along the *x* axis in the central region of the two magnets.

Fig. 2 depicts the simulated two-dimensional (2D) magnetic field magnitudes of the tilted Proteus magnet in the Y-Z, X-Y, and X-Z planes. The magnetic field has contributions from $B_x$, $B_y$, and $B_z$, but in all cases $B_z$ dominates. The proposed gradient strength was selected on the basis of ability to observe flow rates within an average velocity range of 1-5 cm/s with echo times below 1 ms. Fig. 3(a) is the 1D profile of the magnetic magnitude field along the central axis of the X-Z plane, taken from Fig. 2(c). The 60 G/cm gradient $G_x$ is observed ± 0.5 cm about the origin in Fig. 3(a). Simulation shows, in the region of the RF probe, that gradient $G_x$ is uniform to within 3 G/cm when displaced 3.45 mm off the central axis in the X-Y plane and within 2 G/cm in the X-Z plane. In order to ensure that the phase-shift measured would be observed in a region of constant gradient, an RF coil with a length of 0.32 cm and inner diameter of 1.0 cm was placed in the centremost region.

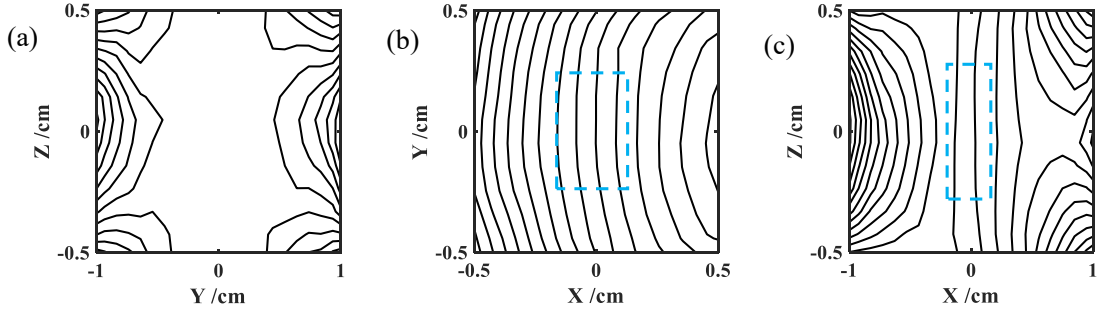

Fig. 2. Simulated 2D magnetic field magnitudes of the tilted Proteus magnet in the central 2D Y-Z, X-Y, and X-Z axis planes. (a) The field plot in the Y-Z plane is largely uniform within the volume of interest. The field contour interval is 7 G. (b) and (c) illustrate the constant gradient in the region of measurement (dashed box). Field contour intervals are 6 and 12 G, respectively.

Fig. 3(b) is the experimental field plot of 1D magnetic field magnitude along the X central axis, Y = 0, Z = 0. As before, the magnetic field has contributions from $B_x$, $B_y$, and $B_z$. The finite size of the field sensor permitted only on axis measurement. From simulation, the region of constant gradient, on axis, was ± 0.5 cm about the origin. The experimental field plot showed the region of constant gradient was reduced to 6 mm compared to simulation. The experimental field plot yields a $G_x$ value, near the origin, of 64 G/cm. The discrepancies in spatial extent and $G_x$ value from simulation are likely due to non-ideal disk magnets. The RF probe was centered about the magnet origin. The Proteus magnet was tuned to a $^1$H frequency of 20.48 MHz. Average velocity measurements of known water flow were performed to confirm the $G_x$ gradient amplitude of 65 G/cm.

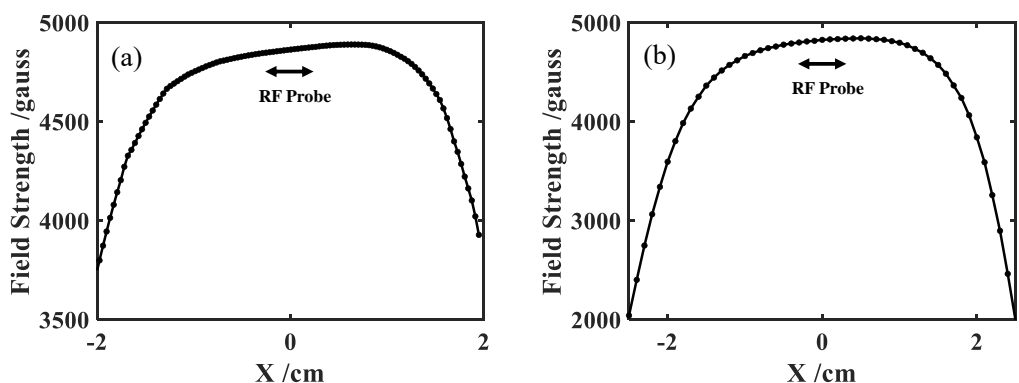

Fig. 3. (a) 1D field magnitude along the central line of the X-Z transverse plane, obtained from Fig. 2(c). (b) Experimental 1D field magnitude measured along X with Y = 0, Z = 0 of the 1° tilted Proteus magnet. The field strength and region of the desired constant gradient are determined coarsely compared to the simulated results, but are, nevertheless, similar. Discrepancies observed in the experimental field compared to the simulated field are likely indicative of imperfections in the disk magnets.

The RF probe was attached to a TecMag (Houston, TX) transcoupler with a λ/4 cable via BNC connectors. The transcoupler was joined to a Tomco Technologies (Stepney, Australia) 250 W RF amplifier and a L3 Nard-MITEQ (Hauppauge, NY) 0.7–200 MHz preamplifier with a Mini-Circuits (Toronto, Ontario) 30 MHz low-band-pass filter. Radio frequency excitation and signal detection were accomplished using a 4-turn solenoidal RF coil driven by a Tomco Technologies 250 W RF amplifier. The system was run by a TecMag LapNMR console.

2.1.2 Flow network

The flow network was identical to the setup previously employed in [50] for time-of-flight flow experiments. In this configuration, a gravity-fed flow from a reservoir suspended several feet above the Proteus magnet was refreshed via a pump from another reservoir at floor level to establish a constant flow through the Proteus magnet. To ensure a constant fluid level in the upper reservoir, and therefore a constant pressure head driving the flow, a submersible pump (Hidom Electric, Shenzhen, China) provided more inflow to the upper reservoir than was flowing through the magnet. An overflow was installed in the upper reservoir to return excess water to the lower reservoir. A Masterflex Variable-Area Flowmeter (Cole-Parmer model # RK-32460-34, Montreal, Canada) was used to control the average flow rate. Flexible Fisherbrand clear PVC tubing (Fisher Scientific Company, Ottawa, Canada) with an inner diameter of 0.8 cm was incorporated throughout the construction of the flow network except for the portion running through the magnet, where a 70 cm length of glass tubing with an ID of 0.67 cm was utilized.

2.2. Basic fluid dynamics model

When a power-law fluid flows in a circular pipe under laminar conditions, the shear stress is proportional to the shear rate raised to the power $n$, where $n$ is the flow behaviour index. Assuming the flow direction in $x$, the constitutive equation can be expressed as [52]

$$\sigma_{xr} = m\dot{\gamma}^n, \qquad (1)$$

where $\sigma_{xr}$ is the shear stress on the radial position $r$, $m$ is the fluid consistency coefficient, $\dot{\gamma}$ is the shear rate and it can be expressed as

$$\dot{\gamma} = \frac{dv(r)}{dr}, \tag{2}$$

where $v(r)$ is the flow velocity at the radial position $r$.

The axial momentum of the fluid in a pipe can be written as

$$0 = -\frac{dp}{dx} + \frac{1}{r}\frac{\partial(r\sigma_{xr})}{\partial r}, \tag{3}$$

where $\frac{dp}{dx} = \frac{\Delta p}{L}$ is the pressure gradient along the pipe. Integrating Eq. (3) with respect to $r$, we can obtain

$$\sigma_{xr} = \frac{r\Delta p}{2L}. \tag{4}$$

Substituting Eqs. (2) and (4) into Eq. (1), we can obtain

$$\frac{r\Delta p}{2L} = m\left(\frac{dv(r)}{dr}\right)^n. \tag{5}$$

Integrating Eq. (5) with respect to $r$, we can obtain the flow velocity profile in a pipe [52]

$$v(r) = \left(\frac{\Delta p R}{2mL}\right)^{\frac{1}{n}} \frac{nR}{n+1}\left(1 - \left(\frac{r}{R}\right)^{\frac{1}{n}+1}\right). \tag{6}$$

The volume flux Q of the pipe flow can be expressed as

$$Q = \int_0^R 2\pi r v(r)\, dr = \frac{\pi n R^3}{1+3n}\left(\frac{\Delta p R}{2mL}\right)^{\frac{1}{n}} = v_{avg}\pi R^2. \tag{7}$$

Eq. (7) shows $v_{avg} = \frac{nR}{1+3n}\left(\frac{\Delta p R}{2mL}\right)^{\frac{1}{n}}$. Substituting $v_{avg}$ into Eq. (6), we obtain

$$v(r) = \frac{3n+1}{n+1} v_{avg} \left(1 - \left(\frac{r}{R}\right)^{\frac{1}{n}+1}\right). \tag{8}$$

For computational convenience, we define $n' = \frac{1}{n} + 1$, and then Eq. (8) can be simplified as

$$v(r) = \frac{n'+2}{n'} v_{avg} \left(1 - \left(\frac{r}{R}\right)^{n'}\right). \tag{9}$$

Eq. (9) shows the maximum flow velocity $v_{max}$ at the centre of the pipe, under laminar conditions, is related to $v_{avg}$, described as

$$v_{max} = \frac{n'+2}{n'} v_{avg}. \tag{10}$$

Different fluids exhibit different $n'$ for pipe flow. For $n' < 2.0$ ($n > 1$), the fluid exhibits shear-thickening behaviour. For $n' = 2.0$ ($n = 1$), the fluid shows Newtonian behaviour. For $n' > 2.0$ ($n < 1$), the fluid shows shear-thinning behaviour. We plot three typical 1D velocity profiles ($n'$ = 1.5, 2.0, and 5.0) of laminar flow at the same $v_{avg}$ = 5 cm/s, in Fig. 4. The velocity profile shape depends on $n'$, and the larger the $n'$, the lower the maximum flow velocity at the same $v_{avg}$. The velocity profile becomes increasingly blunt (more plug-like) as $n'$ increases. Therefore, $n'$ and $v_{avg}$ are the two necessary parameters for determining the flow velocity profile.

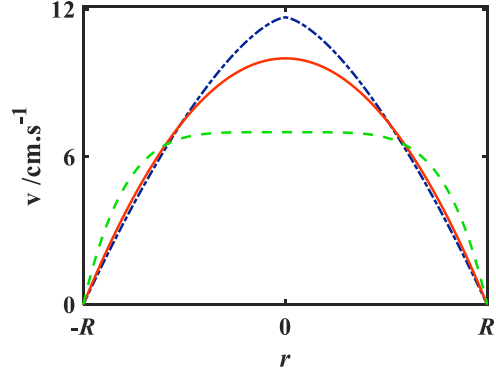

Fig. 4. Three typical 1D velocity profiles of laminar flows at the same $v_{avg}$ = 5 cm/s with $n'$ = 1.5 (----), 2.0 (—), and 5.0 (- - -). The velocity profile shape varies from $n'$, and the larger the $n'$, the lower the maximum flow velocity. The velocity profile becomes increasingly blunt (more plug-like) as $n'$ increases.

2.3. Flow parameter determination from CPMG measurement

2.3.1. Phase-based method

The CPMG MR method is composed of a $90_x^\circ$ pulse followed by a series of $180_y^\circ$ pulses with $2\tau$ time spacing. This measurement can be described as

$$90_x^0 - \left[\tau - 180_y^0 - \tau - echo\right]_j. \quad (11)$$

Each $180_y^\circ$ pulse refocuses the magnetization to generate an echo at the time $t = 2j\tau$, where $j$ denotes the $j$th spin echo during the CPMG measurement. When the CPMG measurement is performed for fluid flow with a constant magnetic field gradient ($G$) in the direction of flow, a phase shift for all odd echoes will occur. For a constant velocity $v_c$, the net phase accumulation $\phi_c$ of any odd echoes can be expressed as [16, 53]

$$\phi_c = \gamma G v_c \tau^2, \quad (12)$$

where $\gamma$ is the gyromagnetic ratio.

For a general steady flow, where flow velocity depends on position, the net phase accumulation $\phi_{odd}$ of odd echoes can be expressed in discrete form

$$\phi_{odd} = \frac{1}{P}\gamma G\tau^2\left(\sum_{p=1}^{P}v_p\right) = \gamma G\tau^2\left(\frac{1}{P}\sum_{p=1}^{P}v_p\right) = \gamma G v_{avg}\tau^2, \tag{13}$$

where $v_{avg}$ is the average velocity. For a general laminar flow in a pipe, we can also calculate the net phase accumulation $\phi_{odd}$ of odd echoes by integration

$$\begin{aligned}\phi_{odd} &= \frac{\iint \phi(r)rdrd\theta}{\iint rdrd\theta} = \frac{\int_0^R \phi(r)rdr}{\int_0^R rdr} = \frac{\int_0^R\left(\gamma Gv(r)\tau^2\right)rdr}{\int_0^R rdr}\\ &= \gamma G\tau^2 \frac{\int_0^R\left(v_{max}\left(1-\frac{r^{n'}}{R^{n'}}\right)\right)rdr}{\int_0^R rdr} = \frac{n'}{n'+2}v_{max}\gamma G\tau^2\\ &= \gamma G v_{avg}\tau^2,\end{aligned} \tag{14}$$

where $R$ is the pipe radius.

It can be seen from Eqs. (13)-(14) that the $\phi_{odd}$ depends on $G$, $v_{avg}$ and $\tau$, and therefore $v_{avg}$ can be determined from $v_{avg}=\phi_{odd}/\gamma G\tau^2$. The $v_{avg}$, determined from net phase accumulation of an echo, suffers from the signal-to-noise ratio (SNR) of the echo measured. To obtain $v_{avg}$ more reliably, multiple first odd echo phase accumulations at different $\tau$ were employed in this paper. The odd echo net phase accumulations at different $\tau^2$ were fitted by Eq. (13), and the slope $k$ can be used to determine $v_{avg}=k/\gamma G$.

### 2.3.2. Magnitude-based method

Under laminar flow conditions, the flow velocity profile is a distribution of velocities, which results in a distribution of accumulated phases at the odd echoes, leading to a change in signal magnitude [51]. Employing odd echo magnitudes is therefore a workable strategy to determine flow parameters. Assuming complete polarization, the odd echo magnitude $M_{odd}$ detected with a flow-directed gradient can be expressed as

$$M_{odd} = M_0 M_R M_\phi, \tag{15}$$

where $M_0$ is the equilibrium magnetization value, which depends on the detected fluid type and quantity. $M_R$ is the normalized magnitude caused by the decay of NMR signal through the so-called spin-spin relaxation process, which has a time constant $T_2$ and depends on the relaxation property of the detected fluid. $M_\phi$ is the normalized magnitude resulting from the phase accumulation, related to the velocity distribution. $M_R$ and $M_\phi$ are not magnitudes but factors reducing magnitude $M_0$ in a strict sense. $M_0$ and $M_R$ are independent of the fluid field, and thus $M_\phi$ can be obtained from dividing the acquired magnitude for stationary solution by the acquired magnitude with flow with the same acquisition parameters.

The normalized signal $S_\phi$ of all odd echoes due to the phase accumulation is the same, and can be expressed as

$$S_\phi = \frac{\iint \exp(-i\phi(r)) r dr d\theta}{\iint ds} = \frac{\int \cos(\phi(r)) r dr}{\int r dr} - i \frac{\int \sin(\phi(r)) r dr}{\int r dr}, \tag{16}$$

where $i$ is the imaginary unit, and $ds = r dr d\theta$ is differential of cross-sectional area. From Eq. (16), the normalized real signal $S_{Re} = \frac{\int \cos(\phi(r)) r dr}{\int r dr}$ and the normalized imaginary signal $S_{Im} = -\frac{\int \sin(\phi(r)) r dr}{\int r dr}$ due to the phase accumulation for all odd echoes. For a circular pipe with a radius of $R$, they can be modified

$$\begin{aligned}
S_{Re} &= \frac{\int_0^R \cos(\phi(r)) r dr}{\int_0^R r dr} = \frac{\int_0^R \cos\left(X\left(1 - \frac{r^{n'}}{R^{n'}}\right)\right) r dr}{\int_0^R r dr} \\
&= \frac{e^{Xi}(-i)^{\frac{2}{n'}}\left(\Gamma\left(\frac{2}{n'}\right) - \Gamma\left(\frac{2}{n'}, Xi\right)\right) + e^{-Xi} i^{\frac{2}{n'}}\left(\Gamma\left(\frac{2}{n'}\right) - \Gamma\left(\frac{2}{n'}, -Xi\right)\right)}{n' X^{\frac{2}{n'}}},
\end{aligned} \tag{17}$$

and

$$S_{\text{Im}} = -\frac{\int_0^R \sin(\phi(r)) r \, dr}{\int_0^R r \, dr} = -\frac{\int_0^R \sin\left(X\left(1 - \frac{r^{n'}}{R^{n'}}\right)\right) r \, dr}{\int_0^R r \, dr} \qquad (18)$$

$$= i \frac{e^{Xi}(-i)^{\frac{2}{n'}}\left(\Gamma\left(\frac{2}{n'}\right) - \Gamma\left(\frac{2}{n'}, Xi\right)\right) - e^{-Xi} i^{\frac{2}{n'}} \left(\Gamma\left(\frac{2}{n'}\right) - \Gamma\left(\frac{2}{n'}, -Xi\right)\right)}{n' X^{\frac{2}{n'}}},$$

where $X = \frac{n'+2}{n'} \gamma G v_{\text{avg}} \tau^2$ and $\Gamma(a, x) = \int_x^\infty w^{a-1} e^{-w} dw$. Detailed derivations of Eqs. (17) and (18) are given in Appendix A. The normalized magnitude $M_\phi$ of odd echoes due to the phase accumulation can be calculated from

$$M_\phi = \sqrt{(S_{\text{Re}})^2 + (S_{\text{Im}})^2} = \frac{2}{n' X^{\frac{2}{n'}}} \sqrt{\left(\Gamma\left(\frac{2}{n'}\right) - \Gamma\left(\frac{2}{n'}, Xi\right)\right)\left(\Gamma\left(\frac{2}{n'}\right) - \Gamma\left(\frac{2}{n'}, -Xi\right)\right)}. \qquad (19)$$

It can be seen from Eq. (19) that $M_\phi$ is not only related to instrument and acquisition parameters, $G$ and $\tau$, but also to laminar flow parameters, $n'$ and $v_{\text{avg}}$. Based on Eqs. (14) and (19), we present several schemes for determining the laminar flow parameters, as follow:

Scheme 1: We use only one first odd echo to calculate laminar flow parameters. The $v_{\text{avg}}$ is determined from the echo net phase accumulation using Eq. (14), and then $n'$ is solved by Eq. (19) with the calculated $v_{\text{avg}}$. This scheme, involving one odd echo data point, suffers from noise, and thus the results have a poor reliability for realistic flow measurements.

Scheme 2: Based on the magnitude data of first odd echoes at different $\tau$, the $v_{\text{avg}}$ and $n'$ are directly fitted by Eq. (19). Due to the complexity of Eq. (19), insufficient data detected might affect the fitting accuracy for this scheme.

Scheme 3: Based on the net phase accumulation of first odd echoes at different $\tau$, the $v_{\text{avg}}$ is fitted by Eq. (14). Subsequently, the $n'$ is fitted by Eq. (19) based on the magnitude data of first odd echoes at different $\tau$ and the fitted $v_{\text{avg}}$.

Schemes 1-3 all employ first odd echo signals detected with CPMG measurement with a flow-directed constant $G$, to solve for the laminar flow parameters, $n'$ and $v_{avg}$. We note it would be feasible to use $v_{avg}$ determined from a known volumetric flow rate and pipe diameter as an alternative to the $v_{avg}$ determined by the net phase accumulation of odd echoes in Schemes 1 and 3. We note in scheme 1-3 above the use of the first odd echo in a CPMG acquisition means that subsequent data points result from additional first odd echoes in new CPMG acquisitions with different $\tau$.

2.3.3. Velocity spectrum method

The complex signal $S(q)$ of all odd echoes in a CPMG measurement after removing the diffusion effect can be expressed as [21, 54]

$$S(q) = \int_{-\infty}^{+\infty} p(v)\exp(-iqv)dv, \qquad (20)$$

where $q = \gamma G \tau^2$, and $p(v)$ is the velocity spectrum. Eq. (20) shows that $S(q)$ is the Fourier transformation of $p(v)$ with respect to $v$. $p(v)$ can therefore be determined by the inverse Fourier transformation of $S(q)$ with respect to $q$, described as

$$p(v) = \int_{-\infty}^{+\infty} S(q)\exp(iqv)dq. \qquad (21)$$

For an acquisition system with a constant magnetic field gradient, we can only change $\tau$ during the measurement. Note that this method does not involve the use of phase-, frequency-, and motion- encoding magnetic resonance gradients. When the magnetic field gradient is parallel to the flow direction, $q$ is a positive number, and thus Eq. (21) can be modified as

$$p(v) = \int_{0}^{+\infty} S(q)\exp(iqv)dq. \qquad (22)$$

To meet the uniform sampling requirement of $q$ with a Fourier transformation, we increase $\tau^2$ with a constant step size. The Field of Flow (FOF) is determined by $2\pi/\Delta q$, where $\Delta q = \gamma G \Delta(\tau^2)$ is the step size of $q$. Since FOF should be no less than the maximum velocity of flow, a short step size of $\tau^2$ is required. To obtain a velocity spectrum with an adequate resolution, a large number of different $\tau^2$ values may be indicated. A compromise may be required for each measurement to constrain the total measurement time. Velocities greater than the maximum velocity in the flow field should ideally have zero amplitude in the velocity spectrum. One can therefore determine the maximum velocity based on the break point in the velocity spectrum. Combined with the average velocity from the net phase accumulation of odd echoes or volumetric flow rate and pipe diameter, $n'$ can be solved for with a known maximum velocity.

For a Poiseuille flow, one can directly use the velocity spectrum to calculate the flow profile by [20, 21]

$$r^2(v) = R^2 \left[ 1 - \int_{v_{\min}}^{v} p(v) dv \right], \tag{23}$$

where $r(v)$ is the radial position associated with a flow velocity, and $v_{\min}$ is the minimum velocity at $r = R$.

## 3. NUMERICAL SIMULATIONS AND ANALYSES

To assess the presented methods in Section 2.3 for determining the flow parameters in a circular pipe, a few numerical simulation tests were performed. Owing to the use of the normalized data during the simulations, we can replace the simulations on the whole circular pipe with those on a circular cross-section. The cross-section was discretized using a 500 × 500 grid. A diagram of the discretized cross-section via a 10 × 10 grid is shown in Fig. 5. We note a larger grid (> $500^2$) did not show appreciable changes for the normalized signal simulated. The intersections on the grid in the circular cross-section were considered during the simulations. The velocity at each intersection can be calculated based on the flow velocity profile (Eq. 9). The normalized real and imaginary signal due to a phase accumulation, for odd echoes during CPMG measurement, can be written in the discrete form

$$S_{\text{Re}} = \frac{1}{N} \sum_{i=1}^{N} \cos(\gamma G v_i \tau^2), \tag{24}$$

and

$$S_{\text{Im}} = -\frac{1}{N} \sum_{i=1}^{N} \sin(\gamma G v_i \tau^2), \tag{25}$$

where $N$ is the number of intersections in the circular cross-section, and $v_i$ is the velocity at the $i$th intersection. Therefore, the net phase accumulation $\phi_{odd}$ of odd echoes can be calculated by

$$\phi_{odd} = \arctan\left(-\frac{S_{\text{Im}}}{S_{\text{Re}}}\right) = \arctan\left[\frac{\sum_{i=1}^{N} \sin(\gamma G v_i \tau^2)}{\sum_{i=1}^{N} \cos(\gamma G v_i \tau^2)}\right]. \tag{26}$$

The normalized magnitude $M_\phi$ of odd echoes due to the phase accumulation can be determined from

$$M_\phi = \sqrt{(S_{\text{Re}})^2 + (S_{\text{Im}})^2} = \frac{1}{N} \sqrt{\left[\sum_{i=1}^{N} \cos(\gamma G v_i \tau^2)\right]^2 + \left[\sum_{i=1}^{N} \sin(\gamma G v_i \tau^2)\right]^2}. \tag{27}$$

The radius of the cross-section was set to 1 cm and the magnetic field gradient was set to 65 G/cm during the simulations. To match the experimental data, Gaussian noise with a SNR of 50 was added to the real and imaginary signal. We note that the effects of $B_1$ inhomogeneity, resonance offset, and pulse imperfection are not considered in the simulations. We note that all simulations employed first odd echoes in model CPMG echo trains. We verified the effectiveness of the three methods in Section 2.3 to determine the laminar flow parameters via processing the simulated data, as follows.

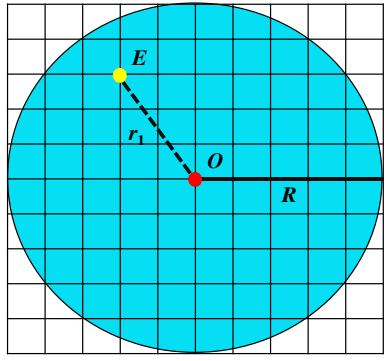

Fig. 5. Diagram of the pipe cross-section discretized via a 10 × 10 grid. We assumed all fluids were positioned on the intersections of the grid in the circular cross-section, and there is nothing inside the single squares. We can calculate the flow velocity at each intersection, for example, $v_E = \dfrac{n'+2}{n'} v_{avg} \left(1 - \dfrac{r_1^{n'}}{R^{n'}}\right)$ at the intersection $E$, where $r_1$ is the distance of the intersection $E$ from the centre $O$.

3.1. Phase-based method verification

The normalized real and imaginary signals of first odd echoes for the three types of laminar flows shown in Fig. 4, at seven different $\tau$ ranging from 100 to 400 μs with a step size of 50 μs, were calculated based on Eqs. (24)-(25). After adding noise, the net phase accumulations can be determined by Eq. (26). Fig. 6 shows the relation of the net phase accumulation $\phi_{odd}$ of first odd echoes to $\tau^2$ for the laminar flows, where $\phi_{odd}$ were the mean of 10 separate simulations and error bars were determined by their standard deviations. From Fig. 6, we can see that the net phase accumulation of first odd echoes, at the same $v_{avg}$, are very close, independent of the flow type.

The simulated data were fitted employing Eq. (13), and the $v_{avg}$ were determined to be $5.01 \pm 0.01$, $4.94 \pm 0.02$, and $5.04 \pm 0.02$ cm/s for Poiseuille flow, shear-thickening flow, and shear-thinning flow, respectively, which are similar to the model $v_{avg} = 5$ cm/s, within 1%. These indicated that the net phase accumulation of first odd echoes can be used to determine the average velocity of any type of laminar flow.

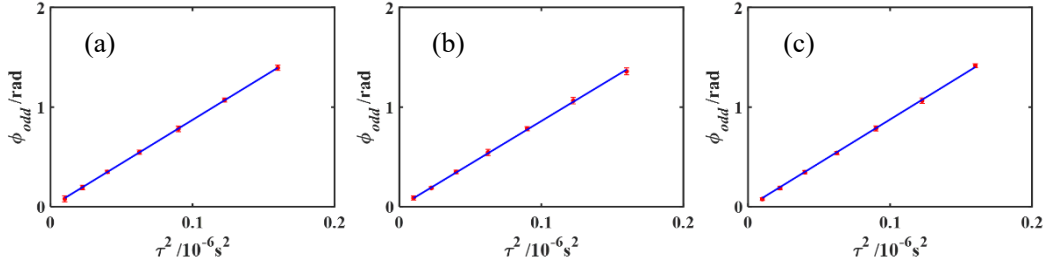

Fig. 6. Fitted results of $\phi_{odd}$ of first odd echoes at different $\tau^2$, for Poiseuille flow (a), shear-thickening flow (b), and shear-thinning flow (c). The fitted relationships between $\phi_{odd}$ of first odd echoes and $\tau^2$ were (a) $\phi_{odd} = (5.01 \pm 0.01)\gamma G \tau^2$, (b) $\phi_{odd} = (4.94 \pm 0.02)\gamma G \tau^2$, and (c) $\phi_{odd} = (5.04 \pm 0.02)\gamma G \tau^2$.

3.2. Magnitude-based method verification

To ensure the accuracy of the magnitude-based method, it is desirable to employ a broader range of echo times. Here the normalized magnitudes $M_\phi$ of first odd echoes due to phase accumulation at 19 different $\tau$ ranging from 100 to 1000 μs with a step size of 50 μs were employed. Fig. 7 shows the relation of $M_\phi$ to $\tau^2$ for laminar flows, where $M_\phi$ are the mean of 10 separate simulations and error bars are determined by their standard deviations. The trends of $M_\phi$ significantly differ with the type of laminar flows at the same $v_{avg}$. Schemes 2 and 3 were both employed to process the simulated magnitude data to obtain laminar flow parameters. For scheme 3, the fitted $v_{avg}$ from the phase-based method in Section 3.1 was used. The fitted results of the two schemes are shown in Fig. 7.

For the Poiseuille flow, $n' = 1.92 \pm 0.03$ and $v_{avg} = 4.93 \pm 0.06$ cm/s by scheme 2, and $n' = 1.98 \pm 0.02$ by scheme 3. For the shear-thickening flow, $n' = 1.54 \pm 0.03$ and $v_{avg} = 5.05 \pm 0.10$ cm/s by scheme 2, and $n' = 1.48 \pm 0.02$ by scheme 3. For the shear-thinning flow, $n' = 4.99 \pm 0.03$ and $v_{avg} = 4.97 \pm 0.03$ cm/s by scheme 2, and $n' = 5.04 \pm 0.03$ by scheme 3. The fitted $n'$ and $v_{avg}$ by scheme 2 agree with the model, within 4% and 1%. Similarly, the fitted $n'$ by scheme 3 are close to the model, within 1%, for the three types of laminar flows. The fitted results indicated that schemes 2 and 3 can both be used to determine the laminar flow parameters by processing the normalized magnitude of odd echoes during CPMG measurement with a flow-directed gradient. The error of the fitted $n'$ by scheme 3 is slightly lower than those by scheme 2, due to there being fewer parameters of scheme 3, revealing that scheme 3 is somewhat superior to scheme 2.

Based on the fitted flow parameters from schemes 2 and 3, the flow velocity profiles were reconstructed and then compared with the model, as shown in Fig. 8. From Fig. 8, we can see that the reconstructed 1D flow velocity profiles for the three laminar flows are very close to the model. The error plots show that the velocity errors in the pipe are less than 0.2 cm/s, revealing the effectiveness of the magnitude-based method to determine flow parameters.

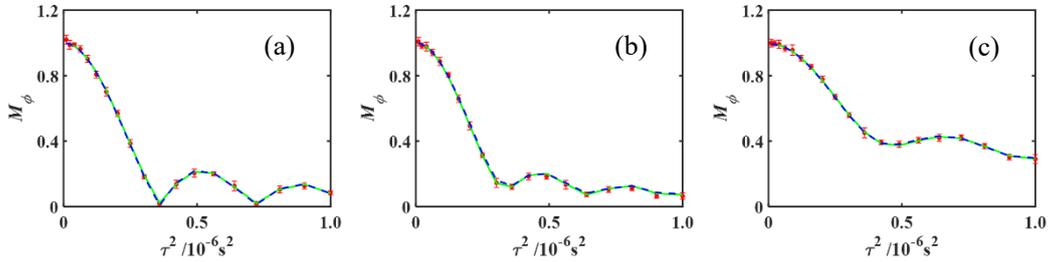

Fig. 7. Fitted results of $M_\phi$ of first odd echoes at different $\tau^2$ using scheme 2 (—) and scheme 3 (----) for Poiseuille flow (a), shear-thickening flow (b), and shear-thinning flow (c). All the $M_\phi$ for different laminar flows decrease with oscillations as $\tau^2$ increases. The trends of $M_\phi$ significantly differ with the type of laminar flows at the same $v_{avg}$. The fitted curves by schemes 2 and 3 are both in agreement with the simulated data.

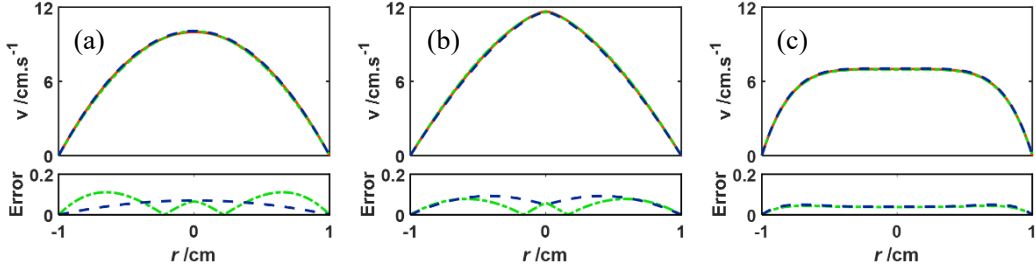

Fig. 8. Comparisons of 1D flow velocity profile models (—) with those reconstructed by scheme 2 (----) and scheme 3 (----) for Poiseuille flow (a), shear-thickening flow (b), and shear-thinning flow (c). The bottom subplots represent the absolute error between the reconstructed profile and model, showing that the velocity errors for any laminar flows in the pipe are less than 0.2 cm/s.

3.3. Velocity spectrum method verification

To meet the requirement of the velocity spectrum method, the data must be sampled with a fixed increment of $\tau^2$. During the simulations, the normalized signals at 128 different $\tau^2$ ranging from $6.25 \times 10^{-4}$ to $31.75$ ms$^2$ with a step size of $0.25$ ms$^2$, for the three flows, were calculated and then Gaussian noise was added. The FOF was therefore 14.45 cm/s. Before undertaking Fourier transformation of simulated data, the exponential filtering method [55] was employed to improve the resolution of the velocity spectrum. The velocity spectra for Poiseuille flow, shear-thickening flow, and shear-thinning flow are shown in Fig. 9. The increase in amplitude at the maximum displayed velocity in the real velocity spectra is an artifact.

It can be seen from Fig. 9 that the characteristics of the velocity spectrum vary depending upon the laminar flow type. Based on their characteristics, the flow type can be identified qualitatively. More importantly, the real, imaginary, and magnitude velocity spectra all have a break point at the same $v$, and the $v_{max}$ of the laminar flows can be determined from the break point. The discontinuity is most readily observed for Poiseuille flow and shear-thinning flow, panels (a) and (c) in Fig. 9. The elevated baseline in panels (a)-(c) of Fig. 9 is due to Fourier transformation of solely +$q$ data points. The $v_{max}$ of the three flows were 10.05 ± 0.11 cm/s, 11.63 ± 0.11 cm/s, and 7.11 ± 0.11 cm/s, respectively. Combined with their $v_{avg}$ from net phase accumulation in Section 3.1, their $n'$ were determined to be 1.99, 1.48, and 4.87, respectively, by Eq. (10). Their calculated $n'$ values are similar to the models, within 3%.

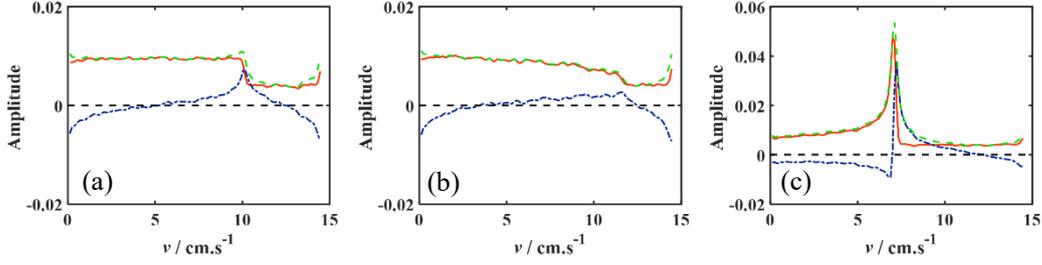

Fig. 9. Real (—), imaginary (----), and magnitude (----) velocity spectra obtained by Fourier transformation of the simulated signals of odd echoes for Poiseuille flow (a), shear-thickening flow (b), and shear-thinning flow (c). Based on the break point in any of the real, imaginary, and magnitude velocity spectra, the maximum velocities of the three flows were 10.05 ± 0.11 cm/s, 11.63 ± 0.11 cm/s, and 7.11 ± 0.11 cm/s, respectively.

For Poiseuille flow, the 1D flow velocity profile was reconstructed by Eq. (23) based on the real velocity spectrum, as shown in Fig. 10. The reconstructed velocity profile coincides with the model, and the absolute error is no more than 0.1 cm/s, which verifies the feasibility of the velocity spectrum method to reconstruct Poiseuille flow profile.

This paper emphasizes laminar flow characterization by simple numerical fitting methods, rather than the Fourier transformation method. We focused on the experimental verifications of the phase-based method and the magnitude-based method and did not perform experiments to validate the velocity spectrum method. In future, we will experimentally examine the velocity spectrum method.

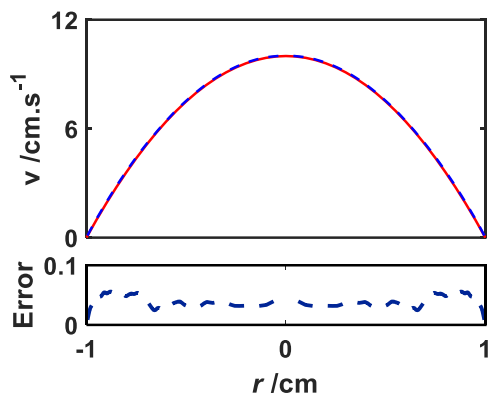

Fig. 10. Comparison of Poiseuille flow velocity profile model (—) with that reconstructed by the velocity spectrum (----). The bottom subplot represents the absolute error between the reconstructed profile and model, showing that the velocity errors for the Poiseuille flow in the pipe are less than 0.1 cm/s.

## 4. EXPERIMENTS

Two solutions, one Newtonian fluid and one shear-thinning fluid, were prepared for flow experiments. Distilled water and glycerol were mixed in a ratio of 6:1 to prepare the Newtonian fluid. Xanthan gum solution, an example of a shear-thinning fluid [23, 56], was prepared in concentration of 0.42 wt% using distilled water. For xanthan gum, complete dissolution was achieved by stirring for 10 h using a low gear mixer (Mastercraft, Toronto, Canada). The two solutions were doped with 0.33 wt% copper sulfate to reduce their $T_1$ lifetimes to ensure the measured fluid was completely polarized. $T_1$ lifetimes of the glycerol/distilled water solution and the xanthan gum solution were 42 ms and 39 ms.

Glycerol/distilled water solution flow experiments were performed at flow rates of $40 \pm 1$ mL/min and $78 \pm 1$ mL/min to produce average velocities of $1.89 \pm 0.05$ cm/s and $3.69 \pm 0.05$ cm/s. Reynolds numbers were 82 and 160 for the two flows, and thus the flows are laminar. Xanthan gum solution flow experiments were performed at flow rates of $35 \pm 1$ mL/min and $66 \pm 1$ mL/min to produce average velocities of $1.65 \pm 0.05$ cm/s and $3.12 \pm 0.05$ cm/s. These are within the laminar flow regime reported is typically observed for Reynolds numbers up to 2000 [57]. All the flow rates were determined from outflow with a measuring cylinder and timer.

The CPMG measurement was employed to measure the two types of flows. Echo CPMG measurement using a single echo time required approximately 2.5 min with a repetition time of 300 ms and 512 averages. The $90_x°$ pulse RF amplitude was set to half the $180_y°$ pulse RF amplitude, and the quadrature echo method [58] was used to set the common $90_x°$ and $180_y°$ pulse durations. Each pulse was 3.2 μs.

The measured magnitude data for flowing fluid were divided by corresponding data collected for a stationary solution with the same measurement parameters, to obtain the normalized magnitude $M_\phi$ of odd echoes due to phase accumulation. Only the phase and $M_\phi$ of the first odd echo were processed employing the phase-based method and the magnitude-based method presented in Section 2.3. This avoids any possible effects due to the pulse imperfection and resonance offset.

Based on the experimental real and imaginary signals, phase accumulations of the first odd echo at different $\tau^2$, for the two types of flows, were calculated. The phase accumulations $\phi$ were plotted with respect to $\tau^2$, as shown in Fig. 11. Since the experimental data had a system phase $\phi_0$, the phase accumulation $\phi$ of odd echoes can be written as

$$\phi = \phi_0 + \phi_{odd} = \phi_0 + \gamma G v_{avg} \tau^2. \tag{28}$$

Based on Eq. (28), the phase accumulations $\phi$ of the first odd echoes at different $\tau^2$ were fitted for each flow by a linear fitting method, and the fitted results were shown as solid lines in Fig. 11. The fitted $\phi_0 = -0.65 \pm 0.01$ rad and $v_{avg} = 1.88 \pm 0.02$ cm/s for the Poiseuille flow at $v_{avg} = 1.89 \pm 0.05$ cm/s, $\phi_0 = -0.64 \pm 0.01$ rad and $v_{avg} = 3.57 \pm 0.04$ cm/s for the Poiseuille flow at $v_{avg} = 3.69 \pm 0.05$ cm/s, $\phi_0 = -0.58 \pm 0.01$ rad and $v_{avg} = 1.62 \pm 0.02$ cm/s for the shear-thinning flow at $v_{avg} = 1.65 \pm 0.05$ cm/s, and $\phi_0 = -0.68 \pm 0.01$ rad and $v_{avg} = 3.15 \pm 0.02$ cm/s for the shear-thinning flow at $v_{avg} = 3.12 \pm 0.05$ cm/s.

Fig. 11 shows that the fitted phase accumulations of the first odd echoes agree with the measured phase accumulation, indicating the reliability of the fitted parameters. The fitted $v_{avg}$ is similar to the $v_{avg}$ from flow rate data of each flow, within 3%. The processed results of experimental data reveal that the phase-based method is feasible and practical in determining the average velocity of laminar flow.

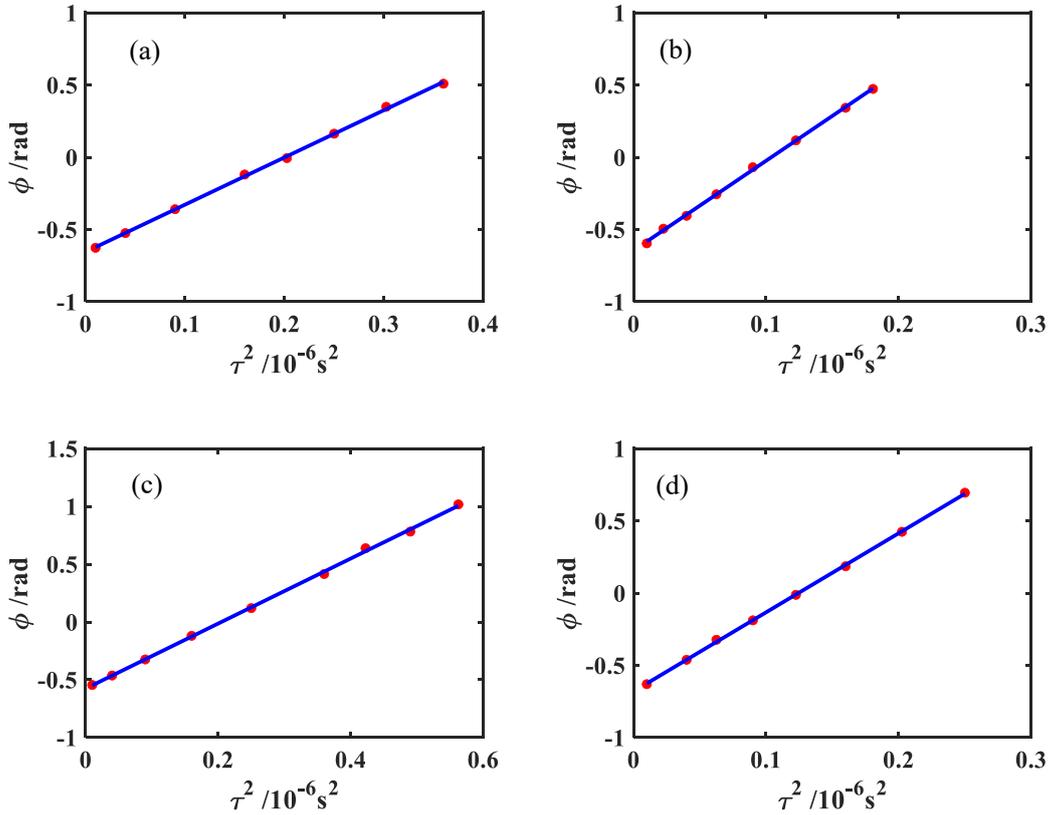

Fig. 11. Processed results of the phase-based method for the glycerol/distilled water flows at $v_{avg} = 1.89 \pm 0.05$ cm/s (a) and $3.69 \pm 0.05$ cm/s (b) and for the xanthan gum solution flows at $v_{avg} = 1.65 \pm 0.05$ cm/s (c) and $3.12 \pm 0.05$ cm/s (d). Red dots show the calculated phase accumulation data of the first odd echo, and the solid line shows the fitted results based on Eq. (28). The fitted $\phi_0 = -0.65 \pm 0.01$ rad and $v_{avg} = 1.88 \pm$

0.02 cm/s for (a), $\phi_0 = -0.64 \pm 0.01$ rad and $v_{avg} = 3.57 \pm 0.04$ cm/s for (b), $\phi_0 = -0.58 \pm 0.01$ rad and $v_{avg} = 1.62 \pm 0.02$ cm/s for (c), and $\phi_0 = -0.68 \pm 0.01$ rad and $v_{avg} = 3.15 \pm 0.02$ cm/s for (d).

The normalized magnitude $M_\phi$ of the first odd echo at different $\tau^2$, for the two types of flows, is displayed as red dots in Fig. 12. After $v_{avg}$ was determined from the phase-based method, scheme 3 was employed to process the experimental data to obtain the flow parameter $n'$. The fitted $M_\phi$ of different flows with respect to $\tau^2$ is shown as solid lines in Fig. 12.

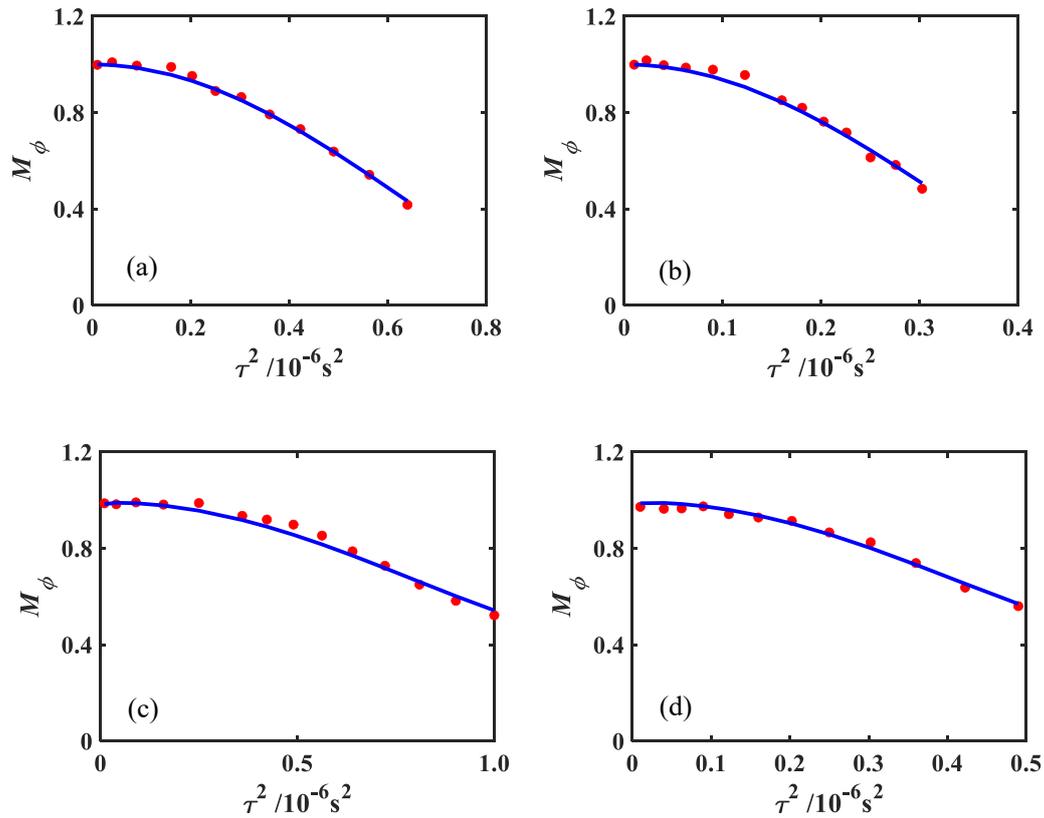

Fig. 12. Processed results of the magnitude-based method (scheme 3) for the glycerol/distilled water flows at $v_{avg} = 1.89 \pm 0.05$ cm/s (a) and $3.69 \pm 0.05$ cm/s (b) and for the xanthan gum solution flows at $v_{avg} = 1.65 \pm 0.05$ cm/s (c) and $3.12 \pm 0.05$

cm/s (d). Red dots show the $M_\phi$ data of the first odd echo, and the solid line shows the fitted results based on Eq. (19). The fitted $n' = 2.11 \pm 0.06$ cm/s for (a), $n' = 1.97 \pm 0.09$ cm/s for (b), $n' = 5.38 \pm 0.19$ cm/s for (c), and $n' = 5.37 \pm 0.17$ cm/s for (d).

The fitted $n'$ were $2.11 \pm 0.06$ and $1.97 \pm 0.09$ for the glycerol/distilled water flows at $v_{avg} = 1.89 \pm 0.05$ cm/s and $v_{avg} = 3.69 \pm 0.05$ cm/s. The fitted $n' = 5.38 \pm 0.19$ and $n' = 5.37 \pm 0.17$ for the xanthan gum solution flows at $v_{avg} = 1.65 \pm 0.05$ cm/s and $3.12 \pm 0.05$ cm/s. A comparison of the fitted $n'$ for the glycerol/distilled water flows with theoretical $n'$ of Poiseuille flow shows that the fitted $n'$ is very close to the theoretical $n'$, within 6%, revealing the effectiveness and practicality of scheme 3 for laminar flows. Both fitted $n'$ are more than 2.0 for the xanthan gum solution flow. These results confirm that the flows are shear-thinning flows, which is as anticipated. The flow behaviour index $n = 0.23$ was determined from the fitted $n'$ at two flow velocities for the xanthan gum solution flow. The calculated $n$ is very similar to that from Blythe *et al.* [22] for similar solution concentrations under laminar conditions, verifying the reliability of scheme 3 for non-Newtonian flows.

Based on the fitted flow parameters $n'$ and $v_{avg}$ by the phase-based method and the magnitude-based method, flow velocity profiles were reconstructed, as shown in Fig. 13. For Poiseuille flow, the flow velocity profiles can be predicted by flow rates due to known $n' = 2.0$. The theoretical predictions are shown as solid lines in Fig. 13 (a), where the bottom subplot represents the absolute error between the reconstructed profile and model. These results show that the velocity errors between reconstructed profiles and theoretical predictions in the pipe are less than 0.2 cm/s, which indicates the reliable accuracy of the phase-based method and the magnitude-based method.

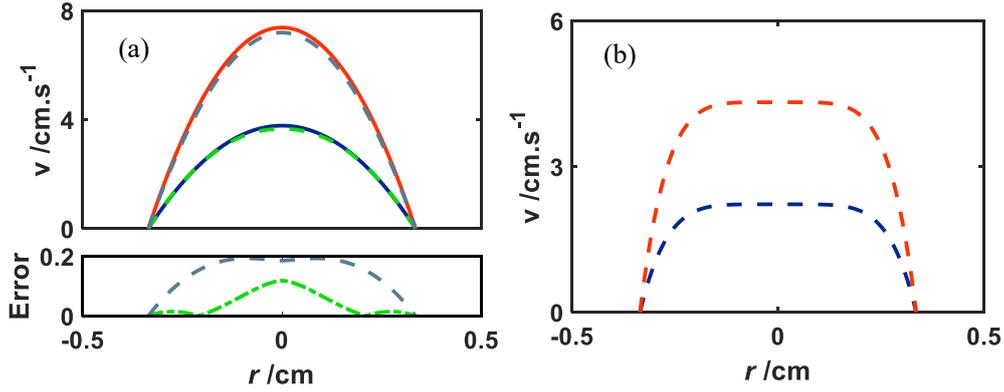

Fig. 13. (a) Comparisons of 1D flow velocity profiles for the glycerol/distilled water flows from the magnitude-based method (dashed line) with the theoretical prediction (solid line), where the bottom subplot represents the absolute error between the reconstructed profile and theoretical profile, showing that the velocity errors between reconstructed profiles and theoretical predictions in the pipe are less than 0.2 cm/s. (b) 1D flow velocity profiles for the xanthan gum solution flows based on the fitted $n'$ and $v_{avg}$.

## 5. CONCLUSIONS AND FUTURE WORK

In this work, a palm-sized Proteus permanent magnet, with a constant magnetic field gradient, was designed to be used for the measurement of laminar flows. The Proteus magnet consists of two portable disk magnets tilted at an angle of 1° from an edge separation of 1.4 cm. Furthermore, we proposed a phase-based method, a magnitude-based method, and a velocity spectrum method to characterize laminar flow in a pipe, including average velocity and velocity profile, from the CPMG measurement. The proposed methods were verified by simulations and flow experiments on the designed magnet. The following conclusions can be drawn:

(1) A phase shift occurs on the odd echoes during CPMG measurement for flow measurement with a flow-directed constant gradient. The phase accumulation is related to gradient, echo time, and average velocity. The phase-based method employs multiple first odd echo phase accumulations at different echo times to fit the average velocity of flow.

(2) The normalized magnitude $M_\phi$ of odd echoes, due to phase accumulation, was derived, dependent on gradient, echo time, average velocity, and the flow behaviour index. The magnitude-based method obtains average velocity and flow behaviour index based on the fitting by $M_\phi$ of first odd echoes at different echo times. With a modest number of first odd echo data points at different echo times, we can obtain the average velocity from the phase-based method, and then fitted flow behaviour index, with known average velocity, by the $M_\phi$.

(3) The velocity spectrum method is based on a Fourier transformation approach. Due to the fixed gradient of the Proteus magnet, this method requires change of $\tau$ during the measurement. The maximum flow velocity can be determined based on the break point in the velocity spectrum. Combined with the average velocity from the phase-based method, the flow behaviour index can be deduced, and in turn the flow profile is determined.

CPMG measurement on low-field MR equipment with a flow-oriented gradient can be directly used for the determination of flow velocity profile based on the proposed methods. Our methods in this paper are developed to process data with complete polarization. The flow measurement, based on our equipment, requires a short $T_1$ of measured fluid, usually ensured using a contrast agent, to make the detected fluid completely polarized. In future work we will consider new magnet designs amenable to pre-polarization which will permit incorporation of incomplete sample magnetization into the flow profile analysis. We will also work to employ additional odd echoes, and even echoes, in CPMG data acquisitions for flow profile measurement.

**ACKNOWLEDGMENTS**

B.J. Balcom thanks NSERC of Canada for a Discovery grant and the Canada Chairs program for a Research Chair in MRI of Materials. B. Newling thanks NSERC of Canada for a Discovery grant (2017-04564). This study was funded by an NSERC Discovery grant (grant number 2015-6122) and CRC grant (grant number 950-230894) held by B.J. Balcom.

**DATA AVAILABILITY**

The data that support the findings of this study are available from the corresponding author upon reasonable request.

# APPENDIX A

## A.1. Derivation of the normalized real signal $S_{\text{Re}}$ of odd echoes due to the phase accumulation

The normalized real signal $S_{\text{Re}}$ of odd echoes due to the phase accumulation can be expressed as

$$S_{\text{Re}} = \frac{\int \cos(\phi(r)) r dr}{\int r dr} = \frac{\int_0^R \cos\left(X\left(1 - \frac{r^{n'}}{R^{n'}}\right)\right) r dr}{\int_0^R r dr}, \quad (A.1)$$

where $X = \frac{n'+2}{n'} \gamma G v_{avg} \tau^2$.

The term $\cos\left(X\left(1 - \frac{r^{n'}}{R^{n'}}\right)\right)$ can be rewritten as

$$\cos\left(X\left(1 - \frac{r^{n'}}{R^{n'}}\right)\right) = \frac{1}{2}\left[\exp\left(iX\left(1 - \left(\frac{r}{R}\right)^{n'}\right)\right) + \exp\left(-iX\left(1 - \left(\frac{r}{R}\right)^{n'}\right)\right)\right]$$

$$= \frac{1}{2}\left[e^{Xi} \exp\left(\left((-Xi)^{\frac{1}{n'}} \frac{r}{R}\right)^{n'}\right) + e^{-Xi} \exp\left(\left((Xi)^{\frac{1}{n'}} \frac{r}{R}\right)^{n'}\right)\right]. \quad (A.2)$$

Thus,

$$\int_0^R \cos\left(X\left(1 - \frac{r^{n'}}{R^{n'}}\right)\right) r dr = \frac{1}{2}\int_0^R \left[e^{Xi} \exp\left(\left((-Xi)^{\frac{1}{n'}} \frac{r}{R}\right)^{n'}\right) r + e^{-Xi} \exp\left(\left((Xi)^{\frac{1}{n'}} \frac{r}{R}\right)^{n'}\right) r\right] dr, \quad (A.3)$$

Eq. (A.3) can be regarded as half of the sum of two integrals $\int_0^R e^{Xi} \exp\left(\left((-Xi)^{\frac{1}{n'}} \frac{r}{R}\right)^{n'}\right) r dr$ and $\int_0^R e^{-Xi} \exp\left(\left((Xi)^{\frac{1}{n'}} \frac{r}{R}\right)^{n'}\right) r dr$.

We define $u = \frac{(Xi)^{\frac{1}{n'}} r}{R}$, and then obtain

$$\int_0^R e^{Xi} \exp\left(\left((-Xi)^{\frac{1}{n'}} \frac{r}{R}\right)^{n'}\right) r dr = \int_0^{(Xi)^{\frac{1}{n'}}} e^{Xi} e^{-u^{n'}} \frac{uR^2}{(Xi)^{\frac{2}{n'}}} du = \frac{R^2 e^{Xi}}{(Xi)^{\frac{2}{n'}}} \int_0^{(Xi)^{\frac{1}{n'}}} e^{-u^{n'}} u du. \quad (A.4)$$

Next, we define $w = u^{n'}$, and Eq. (A.4) can be written as

$$\int_0^R e^{Xi} \exp\left(\left((-Xi)^{\frac{1}{n'}}\frac{r}{R}\right)^{n'}\right) r\, dr = \frac{R^2 e^{Xi}}{(Xi)^{\frac{2}{n'}}} \int_0^{Xi} w^{\frac{1}{n'}} e^{-w} \frac{w^{\frac{1}{n'}-1}}{n'} dw = \frac{R^2 e^{Xi}}{n'(Xi)^{\frac{2}{n'}}} \int_0^{Xi} w^{\frac{2}{n'}-1} e^{-w} dw.$$

(A.5)

The Gamma function $\Gamma(a) = \int_0^{\infty} w^{a-1} e^{-w} dw$ and the incomplete Gamma Function $\Gamma(a, x) = \int_x^{\infty} w^{a-1} e^{-w} dw$. Thus, Eq. (A.5) can be written as

$$\int_0^R e^{Xi} \exp\left(\left((-Xi)^{\frac{1}{n'}}\frac{r}{R}\right)^{n'}\right) r\, dr = \frac{R^2 e^{Xi}}{n'(Xi)^{\frac{2}{n'}}} \int_0^{\infty} w^{\frac{2}{n'}-1} e^{-w} dw - \frac{R^2 e^{Xi}}{n'(Xi)^{\frac{2}{n'}}} \int_{Xi}^{\infty} w^{\frac{2}{n'}-1} e^{-w} dw$$

$$= \frac{R^2 e^{Xi}}{n'(Xi)^{\frac{2}{n'}}} \left[\Gamma\left(\frac{2}{n'}\right) - \Gamma\left(\frac{2}{n'}, Xi\right)\right].$$

(A.6)

Similarly, we can obtain

$$\int_0^R e^{-Xi} \exp\left(\left((Xi)^{\frac{1}{n'}}\frac{r}{R}\right)^{n'}\right) r\, dr = \frac{R^2 e^{-Xi}}{n'(-Xi)^{\frac{2}{n'}}} \left[\Gamma\left(\frac{2}{n'}\right) - \Gamma\left(\frac{2}{n'}, -Xi\right)\right].$$

(A.7)

We substitute Eqs. (A.6) and (A.7) into Eq. (A.3), and then Eq. (A.1) can be rewritten as

$$S_{Re} = \frac{\int_0^R \cos\left(X\left(1-\frac{r^{n'}}{R^{n'}}\right)\right) r\, dr}{\int_0^R r\, dr} = \frac{e^{Xi}(-i)^{\frac{2}{n'}}\left[\Gamma\left(\frac{2}{n'}\right) - \Gamma\left(\frac{2}{n'}, Xi\right)\right] + e^{-Xi} i^{\frac{2}{n'}}\left[\Gamma\left(\frac{2}{n'}\right) - \Gamma\left(\frac{2}{n'}, -Xi\right)\right]}{n'X^{\frac{2}{n'}}}.$$

(A.8)

**A.2. Derivation of the normalized imaginary signal $S_{Im}$ of odd echoes due to the phase accumulation**

The normalized imaginary signal $S_{Im}$ of odd echoes due to the phase accumulation can be expressed as

$$S_{Im} = -\frac{\int_0^R \sin(\phi(r)) r\, dr}{\int_0^R r\, dr} = -\frac{\int_0^R \sin\left(X\left(1-\frac{r^{n'}}{R^{n'}}\right)\right) r\, dr}{\int_0^R r\, dr},$$

(A.9)

where $X = \dfrac{n'+2}{n'} \gamma G v_{avg} \tau^2$.

The term $\sin\left(X\left(1 - \dfrac{r^{n'}}{R^{n'}}\right)\right)$ can be rewritten as

$$\sin\left(X\left(1 - \dfrac{r^{n'}}{R^{n'}}\right)\right) = -\dfrac{i}{2}\left[\exp\left(iX\left(1 - \left(\dfrac{r}{R}\right)^{n'}\right)\right) - \exp\left(-iX\left(1 - \left(\dfrac{r}{R}\right)^{n'}\right)\right)\right]$$
$$= -\dfrac{i}{2}\left[e^{Xi}\exp\left(\left((-Xi)^{\frac{1}{n'}}\dfrac{r}{R}\right)^{n'}\right) - e^{-Xi}\exp\left(\left((Xi)^{\frac{1}{n'}}\dfrac{r}{R}\right)^{n'}\right)\right].$$

(A.10)

Thus,

$$\int_0^R \sin\left(X\left(1 - \dfrac{r^{n'}}{R^{n'}}\right)\right) r\, dr = -\dfrac{i}{2}\int_0^R \left[e^{Xi}\exp\left(\left((-Xi)^{\frac{1}{n'}}\dfrac{r}{R}\right)^{n'}\right) r - e^{-Xi}\exp\left(\left((Xi)^{\frac{1}{n'}}\dfrac{r}{R}\right)^{n'}\right) r\right] dr.$$

(A.11)

Eq. (A.11) can be regard as $-\dfrac{i}{2}$ times the difference of two integrals

$\int_0^R e^{Xi}\exp\left(\left((-Xi)^{\frac{1}{n'}}\dfrac{r}{R}\right)^{n'}\right) r\, dr$ and $\int_0^R e^{-Xi}\exp\left(\left((Xi)^{\frac{1}{n'}}\dfrac{r}{R}\right)^{n'}\right) r\, dr$. Eqs. (A.6) and (A.7)

give the two integral expressions. Thus, Eq. (A.9) can be rewritten as

$$S_{Im} = -\dfrac{\int_0^R \sin\left(X\left(1 - \dfrac{r^{n'}}{R^{n'}}\right)\right) r\, dr}{\int_0^R r\, dr} = i\dfrac{e^{Xi}(-i)^{\frac{2}{n'}}\left[\Gamma\left(\dfrac{2}{n'}\right) - \Gamma\left(\dfrac{2}{n'}, Xi\right)\right] - e^{-Xi} i^{\frac{2}{n'}}\left[\Gamma\left(\dfrac{2}{n'}\right) - \Gamma\left(\dfrac{2}{n'}, -Xi\right)\right]}{n' X^{\frac{2}{n'}}}.$$

(A.12)